\documentclass[%
 aip,
 amsmath,amssymb,
reprint,%
Conference Proceedings
]{revtex4-1}

\usepackage{graphicx}% Include figure files
\usepackage{dcolumn}% Align table columns on decimal point
\usepackage{bm}% bold math
\usepackage[utf8]{inputenc}
\usepackage[T1]{fontenc}
\usepackage{mathptmx}

\begin{document}

\preprint{AIP/123-QED}

\title{Avoiding the $H_c=0$ Anomaly using FORC+ (expanded version of paper GG-05, MMM-Intermag 2019)}
% Force line breaks with \\

\author{P. B. Visscher}
\affiliation{ MINT Center, U of Alabama, Tuscaloosa, AL 35487-0209
%\\This line break forced with \textbackslash\textbackslash
}%

\date{\today}% It is always \today, today,

\begin{abstract}
In conventional FORC (First Order Reversal Curve) analysis of a magnetic system, reversible and low-coercivity irreversible materials are treated as being qualitatively different: the FORC distribution shows low-coercivity materials but completely hides reversible (zero-coercivity) ones.  This distinction is artificial – as the coercivity approaches zero, the physical properties of an irreversible material change smoothly into those of a reversible material.  We have developed a method (called FORC+, implemented in free software at \url{http://MagVis.org}) for displaying the reversible properties of a system (a reversible switching-field distribution, R-SFD) together with the irreversible ones (the usual FORC distribution), so that there is no sudden discontinuity in the display when the coercivity becomes zero.
%To illustrate the advantages of this approach, we will consider a simple model system of Preisach hysterons (i.e., elements with rectangular hysteresis loops) with a distribution of coercivities.  We convert this to a reversible system gradually, as though simulating an oxidation or annealing process that degrades the coercivity -- each hysteron's coercivity decreases by the same amount in each time step, until it is zero.  When the coercivities are all zero, the conventional FORC distribution of the system would disappear entirely, but in FORC+ the density simply shifts to the R-SFD.
%When we compute a FORC distribution from a set of measured FORC curves, some information (in particular, all reversible information) is lost. The addition of the R-SFD decreases the amount of information that is lost, but there is still some loss.
We will define a "FORC+ dataset" to include the usual FORC distribution, the R-SFD, the saturation magnetization, and what we will call the "lost hysteron distribution" (LHD) such that {\bf no} information is lost -- the original FORC curves can be {\bf exactly} recovered from the FORC+ dataset.
We also give some examples of the application of FORC+ to real data -- it uses a novel complementary-color display that minimizes the need for smoothing.  In systems which switch suddenly (thus having sharp structures in the FORC distribution) direct display of un-smoothed raw data allows visualization of sharp structures that would be washed out in a conventional smoothed FORC display.

This is an expanded version of paper GG-05, MMM-Intermag 2019, with a discrete derivation of the FORC distribution (Eq. \ref{dist}) and an additional example (Fig. \ref{Hept}).
\end{abstract}

\maketitle

\section{\label{intro}Introduction}
\begin{figure}[htb]
\begin{center}
\includegraphics[width=2.5in]{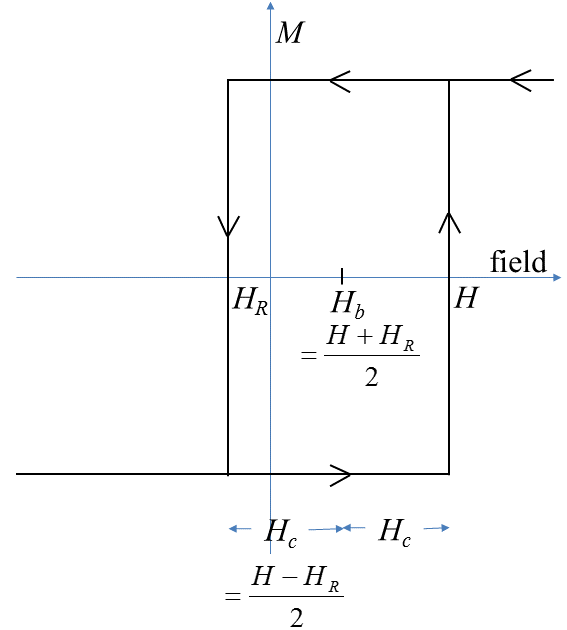}
\caption{\label{hyst} MH loop of a single Preisach hysteron, showing down-switching field $H_R$ and up-switching field $H$, and defining the bias field $H_b$ and the coercivity $H_c$.}
\end{center}
%\label{figure:MI} this causes references to give section number, not figure number!!
%\vspace{-3mm}
\end{figure}

The FORC method\cite{Pike2003,PikeNi2005,Stancu2013,RobertsReview2014} for the characterization of magnetic systems has come into wide use in the past 15 years or so.  It was originally designed for completely irreversible systems, modeled as a collection of Preisach hysterons, each of which has a rectangular hysteresis loop (Fig. \ref{hyst}).
As we lower the field from a large positive saturating value, it switches down at a field usually denoted by $H_R$, (the subscript R stands for "reversal", for reasons that will become apparent later) and as we increase the field it switches back up at a field
$H > H_R$. The Preisach distribution is the density of these hysterons in the $H - H_R$ plane. %(Fig. \ref{prei1}).

\begin{figure}[htb]
\begin{center}
\includegraphics[width=3.0in]{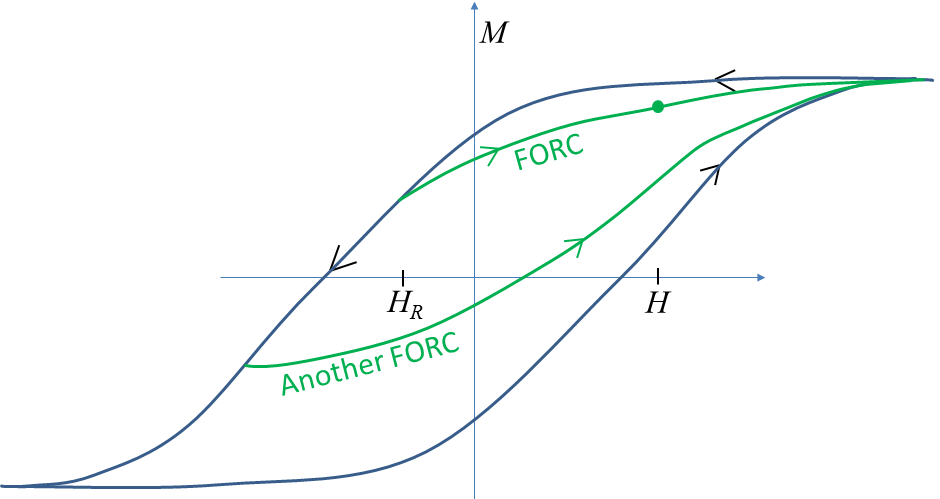}
\caption{\label{FORC} A major hysteresis loop with two FORC curves, with a dot showing the point where $m(H,H_R)$ is defined.}
\end{center}
\end{figure}

The fundamental result behind the FORC idea is that this Preisach distribution can be obtained by measuring "first order reversal curves" -- that is, by saturating the sample in the positive direction, decreasing the field to a reversal field $H_R$ (see Fig. \ref{FORC}), then reversing dH/dt from negative to positive and measuring the magnetization $m(H,H_R)$ as the field increases again past each value $H$.
The distribution of hysterons is then given by
\begin{equation}\label{dist}
\rho(H,H_R)=-\frac{1}{2} \frac{\partial^2 m(H,H_R)}{\partial H \partial H_R}
\end{equation}
It is surprisingly difficult to find a simple derivation of this result in the literature -- we will give a discrete derivation in Sec. \ref{Invert} below.
Note that in a reversible system the moment $m(H,H_R)$ measured at a field $H$ is independent of the history, in particular independent of the reversal field $H_R$.  Thus the derivative (Eq. \ref{dist}) vanishes -- the FORC distribution gives no information about reversible behavior.
In Sec. \ref{SFD}, we will show that the reversible behavior can be characterized by a reversible switching-field distribution (R-SFD), and in Sec. \ref{Invert} we will define a "FORC+ dataset" that includes this reversible information but also enough additional information that the computation of the FORC+ dataset from a set of measured FORC curves is exactly invertible -- we can recover the FORC curves from the FORC+ dataset.  Finally, in Section \ref{examp}, we will show some applications to FORC data on various types of samples.

\section{\label{SFD} Reversible switching field distribution (R-SFD)}
The switching field distribution is most easily defined in a reversible system in which the moment is a function only of the field: $m(H)$.  Then we can think of the system as a superposition of objects ("anhysterons") that switch reversibly at a bias field $H$, and the SFD $S(H)=dm(H)/dH$ gives the total moment $S(H) \Delta H$ of the anhysterons that switch in the field interval $\Delta H$ around $H$.  However, the term "SFD" is also used to refer to $dm/dH$ in a hysteretic system, where $m$ is taken to be the upper branch of the hysteresis loop.  In terms of our notation, these points are the beginnings of FORC curves, where $H=H_R$, so the SFD is
\begin{eqnarray}
\label{dmdH}
S(H) = \frac{dm(H,H)}{dH} = \nonumber \\
 \left( \frac{\partial m(H,H_R)}{\partial H} \right)_{H_R=H} +  \left( \frac{\partial m(H,H_R)}{\partial H_R} \right)_{H_R=H}
\end{eqnarray}
Note that the second term vanishes in a reversible system, and the first term vanishes for an irreversible Preisach hysteron (with nonzero coercivity).  Thus it is natural to define the first term as the "reversible SFD" (R-SFD)
\begin{equation}\label{RSFD}
S_{rev}(H) = \left( \frac{\partial m(H,H_R)}{\partial H} \right)_{H_R=H}
\end{equation}
and the second as the irreversible SFD
\begin{equation}\label{ISFD}
S_{irr}(H) = \left( \frac{\partial m(H,H_R)}{\partial H_R} \right)_{H_R=H}
\end{equation}
We can obtain a measure of the overall reversibility of our system by integrating these SFDs over all $H$ -- the total irreversible moment is
\begin{equation}\label{mirr}
m_{irr} = \int S_{irr}(H) dH
\end{equation}
and similarly the reversible moment is
\begin{equation}\label{mrev}
m_{rev} = \int S_{rev}(H) dH
\end{equation}
and the reversible fraction is
\begin{equation}\label{frev}
f_{rev} = m_{rev} / (m_{irr} + m_{rev}).
\end{equation}
We need not save the I-SFD ({\it i.e.,} $S_{irr}$) separately from the FORC distribution, since it can be obtained by integrating Eq. \ref{dist} with respect to H downward from saturation.  However, the R-SFD is new information, and the FORC+ program displays it.
%An example is shown in Fig. \ref{MovieABC}(a).

\begin{figure}
\includegraphics[width=3.0in]{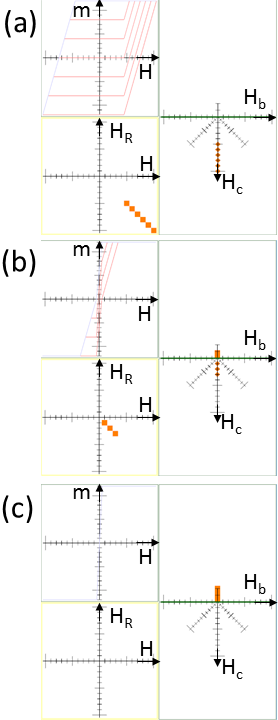}
\caption{\label{MovieABC} (a) The FORC+ display, for a synthetic system of 6 hysterons.  The upper left graph shows the FORC curves; below it is the FORC distribution.  The lower right also shows the FORC distribution, but rotated so the horizontal axis is the bias field $H_b$, and above it is the reversible SFD. The coercivities are assumed to decrease with time (b) until the system is reversible (c).}
\end{figure}

The FORC+ program \cite{FORC+} is driven by a script file, which can either read a magnetometer data file or create a set of synthetic FORC curves for a specified set of Preisach hysterons.
In Fig. \ref{MovieABC}(a) we show the FORC+ display for a synthetic system of six hysterons of different coercivity but zero bias -- this gives a positive Preisach density, which is displayed as an orange color, with primary color values $RGB =(1,0.5,0)$. This particular distribution has no negative density, but negative density is displayed in the complementary color $(0,0.5,1)$.  This is important because if there is noise (a mixture of equal amounts of positive and negative density) this will appear from a distance as a shade of grey, $RGB = (x,x,x)$.  If the grey is tinged with orange, that means there is a small net positive density, and it is not always necessary to do artificial smoothing to see this.
The FORC curves are shown in the upper left, and below them the FORC distribution, with the same $H$ scale so each feature in the distribution appears exactly below the FORC curve that created it.  The FORC distribution is repeated on the lower right, but rotated by $45^\circ$  % or $\textdegree$
so the coercivity axis is vertical (and points downward).  This is done so the R-SFD can be plotted upward as a bar graph, starting from the $H_c=0$ zero-coercivity line.  This ensures a smooth transition as the coercivity of each hysteron decreases (Fig. \ref{MovieABC}(b),(c)) -- it moves upward in the FORC distribution, and when each hysteron reaches $H_c=0$ it reappears at the same field value in the upper right display (R-SFD).  In Fig. \ref{MovieABC}(c), all of the hysterons have become reversible.   The horizontal axis is the bias field $H_b$ in the lower display and $H$ in the upper one, but these are equal along the $H_c=0$ line so they match smoothly.  The display scale (number of orange pixels per emu of magnetic moment) has been chosen to be the same in the FORC distribution and in the bar graph of the R-SFD, so the total number of orange pixels remains the same as the system becomes reversible.  It is common to over-interpret a very weak irreversible FORC distribution in a system that is primarily reversible, because the reversible part is invisible in an ordinary FORC display, and the display program sets the gain very high to normalize the weak signal -- using FORC+, it will be clear that the R-SFD is dominant in such a case.  It is also possible to turn up the gain to make the weaker signal visible (see Fig. \ref{Hept} below), but this can be indicated on the display.  In addition to the graphics window shown in Fig. \ref{MovieABC}, the FORC+ program also opens a text window with additional information about the system (for example, the reversible fraction, Eq. \ref{frev}).

There has been previous work on including reversible effects in the FORC formalism\cite{Pike2003,RevIrrev2008}, some of which was aimed at incorporating the reversible material into the FORC distribution as a singular delta function, and some of which involved a separate plot of the reversible part, but none of these ensured that exactly the same scale was used for the reversible and irreversible displays.

\section{\label{Invert} Invertibility: no information loss}
\begin{figure}
\includegraphics[width=3.5in]{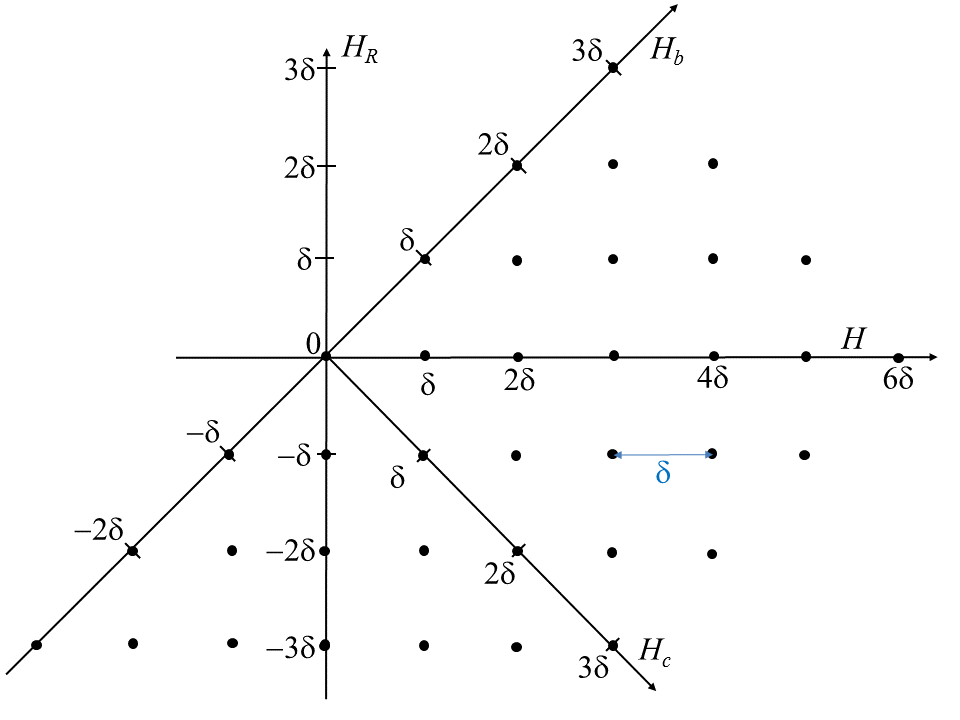}
\caption{\label{dots} The points in the $H$-$H_R$ plane at which $m$ might be measured in a FORC experiment.  Each horizontal string of dots is a FORC curve.}
\end{figure}

In this section we will construct the "FORC+ data set" which preserves all the information in the FORC curve measurement.  An illustrative set of FORC measurements on a regular grid in the $H$, $H_R$ plane is shown in Fig. \ref{dots}.
Each black dot represents a single measurement of the magnetic moment $m(H,H_R)$, for a particular pair of fields ($H$ is the horizontal axis and $H_R$ the vertical one.)
(Of course, no real measurement will give fields on an exactly regular grid.  The FORC+ program operates on an arbitrary grid.  However, it is often nearly regular, and it is very much easier to prove the results we prove below using a regular grid.  The generalization to an arbitrary grid will be published elsewhere.)
Each FORC curve is a horizontal string of black dots, starting at the reversal field $H=H_R$ and ending at some higher field.  In the commonly used PM (Princeton Measurements) FORC protocol, the maximum field is determined by bounds on the bias field $H_b$ or the coercivity $H_c$, so the boundaries of the measured region are lines of constant $H_b$ (perpendicular to the $H_b$ axis, which is also shown) or constant $H_c$.

\begin{figure}
\includegraphics[width=3.5in]{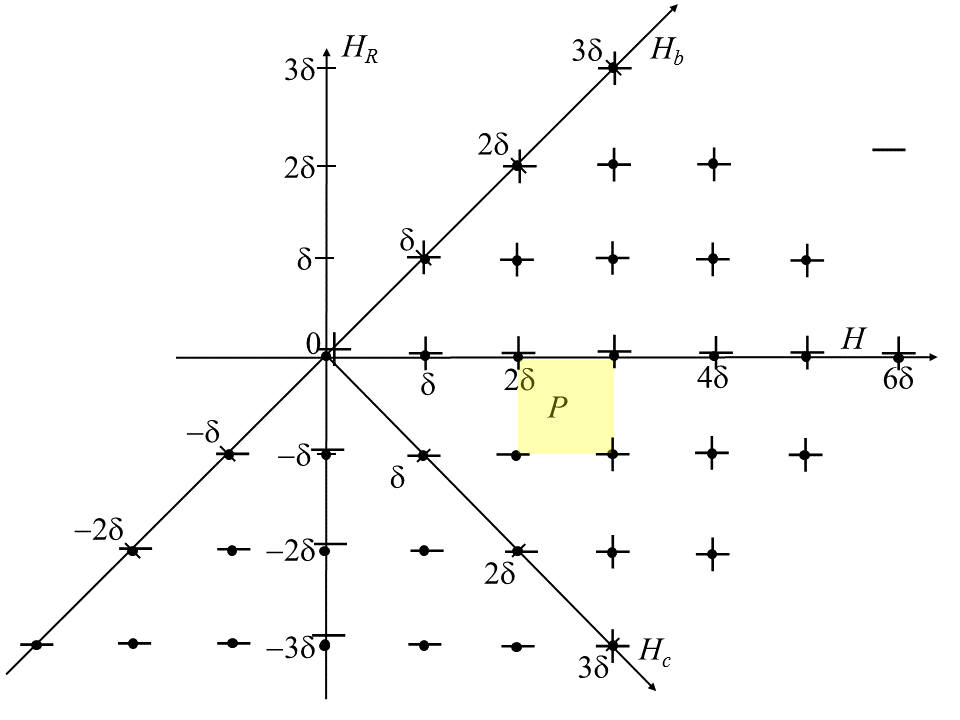}
\caption{\label{Signs} A discrete derivation of Eq. \ref{dist}, relating the FORC distribution to the crossed partial derivative of $m(H,H_R)$. We assume a single Preisach hysteron in the plaquette labeled P.   (It could be at $H=2.5\delta$, $H_R = -0.5\delta$, or anywhere else in this plaquette.) The moment $m(H,H_R)$ has the same magnitude $m_s$ everywhere, and the sign given in the figure.  For reversal field $H_R>-0.5\delta$, the hysteron has not switched down by the beginning of the FORC curve, and a $+$ sign is indicated.  When $H_R>-0.5\delta$, it has switched down so we have
$–$ % fails?
minus signs until $H>2.5\delta$, after which it switches up again.  All plaquettes have either 4 or 2 adjacent $+$ signs, so the second derivative $M_{HRH}$ (defined in the inset to Fig. \ref{wInset}) vanishes, except the one containing P, which has 3 $+$ signs, and
$m_{HRH}  = m_s - m_s - m_s - m_s = -2m_s$, hence $\rho = m_s/\delta^2$, the correct density.}
\end{figure}

Fig. \ref{Signs} shows a discrete derivation of Eq. \ref{dist}, for a single hysteron in the yellow-shaded plaquette labeled $P$.  Because all equations are linear, it follows that it is correct for an arbitrary distribution of hysterons. (Another discrete derivation, not using linearity in this way, has been given previously\cite{Arxiv}.)

\begin{figure}
\includegraphics[width=3.5in]{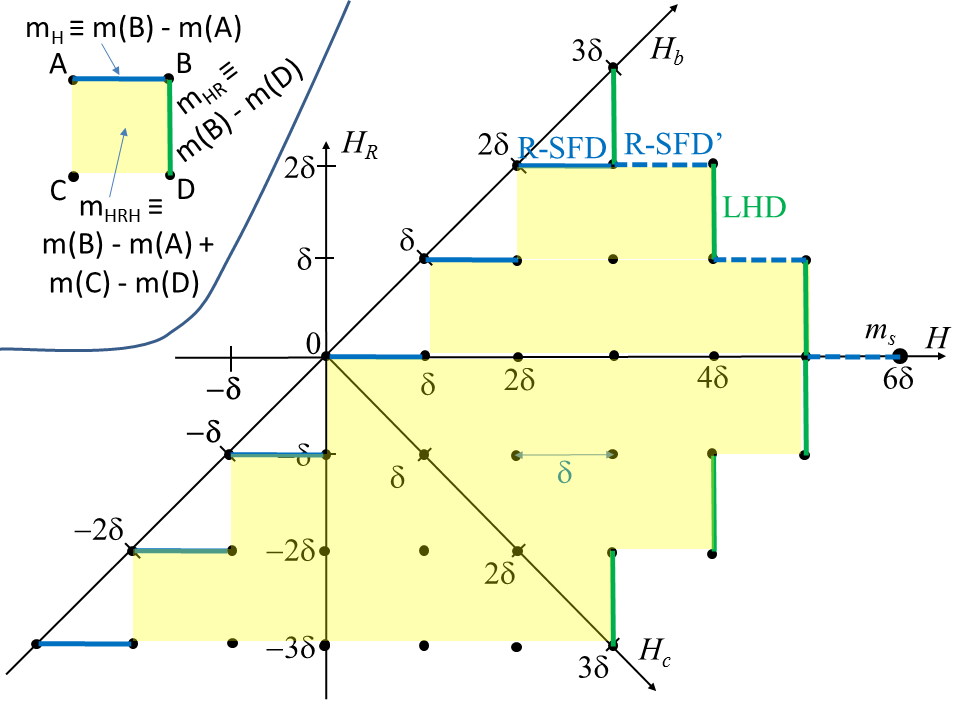}
\caption{\label{wInset} The information included in the FORC+ data set: R-SFD, LHD, $m_s$, and the FORC distribution $m_{HRH}$ (see text).  Inset shows definitions of the vertical and horizontal differences $m_{HR}$ and $m_H$, and the second difference $m_{HRH}$}
\end{figure}

Fig. \ref{wInset} shows the information (the "FORC+ data set") that is necessary to invert the FORC distribution calculation and recover the original data $m(H,H_R)$.
The inset shows a single square plaquette, with four corners $A$, $B$, $C$, and $D$ where the moment $m$ has been measured.  The discrete derivative $dm/dH$ is naturally defined at the blue line segment from $A$ to $B$.  We will denote it (without the factor of $1/\delta$, for convenience) by $m_H = m(B)-m(A)$.  Similarly, $m_{HR}$ is the difference in the vertical direction, and $m_{HRH}$ is the second difference, defined for each plaquette, and related to the FORC density (Eq. \ref{dist}) by $\rho = -\frac{1}{2} m_{HRH} / \delta^2$.
It can be seen that if we know the vertical derivative $m_{HR}$ at the right (along BD) we can compute it at the left using
\begin{equation}\label{integ}
m_{HR}(AC) = m_{HR}(BD) - m_{HRH}
\end{equation}
 -- we are basically integrating the crossed partial derivative with respect to $H$.

Our FORC+ data set will include the Preisach density, i.e. the second derivative $m_{HRH}$, at each of the plaquettes in the yellow-shaded region in Fig. \ref{wInset}. The vertical differences $m_{HR}$ are assumed to go to zero far to the right where $m$ saturates, but will not be zero at the 6 vertical green lines on the right boundary.  In fact, we could calculate them if we knew the Preisach density to the right of the green lines, using Eq. \ref{integ} -- $m_{HR}$ at the green line is basically the total (un-measured) hysteron density to its right, so we will call it the
"lost hysteron distribution" or LHD.  (Even if we measured the FORC until saturation, so there are no lost hysterons, the values of $m_{HR}$ would still be nonzero, due to noise.)  Thus we must store the LHD as part of the FORC+ data set.  From this, we can use Eq. \ref{integ} to calculate the vertical differences $m_{HR}$ everywhere in the yellow region where $m_{HRH}$ is known.

The FORC+ data set must also include the R-SFD (Eq. \ref{RSFD}), giving the horizontal derivatives on the 6 blue segments on the left side of Fig. \ref{wInset}.  We will need horizontal derivatives at the 3 dashed blue lines at the upper right as well, labeled R-SFD'.  We will graph these as part of the R-SFD function -- if there are no lost hysterons above them, the system is completely reversible at these high fields.  In addition to differences, we will need one actual value for the moment, which we take to be at the rightmost point labeled "$m_s$".  Then we alternately add horizontal and vertical differences to recover $m$ at all the points on the upper right boundary.  When we get to the top, we can come back down along the $H_b$ axis, again alternately adding horizontal and vertical differences (recall that we computed the vertical differences everywhere by integrating $m_{HRH}$ from the right).  Now we know $m$ at the top of each column of dots, and can compute them downward along the column using the vertical differences.

Thus we have shown that the FORC+ data set consisting of (1) the FORC distribution $m_{HRH}$, (2) the R-SFD, (3) the lost hysteron distribution LHD, and (4) the saturation moment $m_s$ allows us to exactly recover all the original measurements of $m$.  The figures have been drawn for a particular choice of FORC lengths, but it is not hard to see that the argument works as long as the measured region in the $H$ - $H_R$ plane is convex (concave sections will cause problems).

\section{\label{examp} Examples}
\begin{figure}
\includegraphics[width=3.5in]{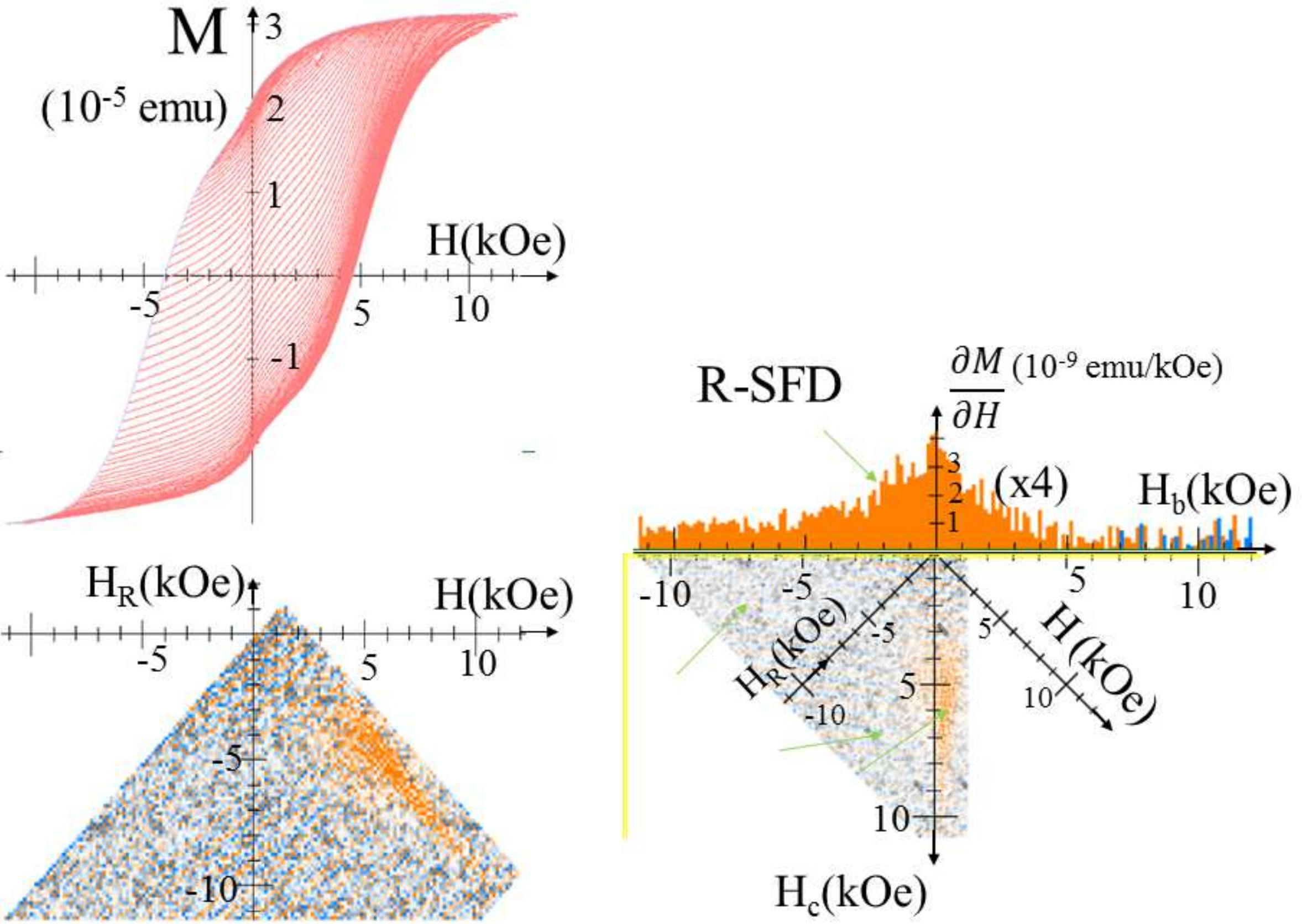}
\caption{\label{Hept} FORC+ display for a patterned CoPt alloy film\cite{heptane}, courtesy of Dr. Allen Owen.}
\end{figure}

Figure \ref{Hept} shows the FORC+ display produced by real data from a Princeton Measurements MicroMag 2900 AGM magnetometer, for a patterned perpendicular film\cite{heptane}.
The FORC distribution (lower left) shows some noise -- to see the orange area (positive density peak) we have turned up the gain so the center of the peak is oversaturated. But there is a clear peak, with a narrow distribution of bias field $H_b$ and a broader distribution of coercivity $H_c$.   Although the reversible fraction is $44\%$, we have multiplied the scale of the reversible SFD at the upper right by 4 so it can be seen more easily.

\begin{figure}
\includegraphics[width=3.5in]{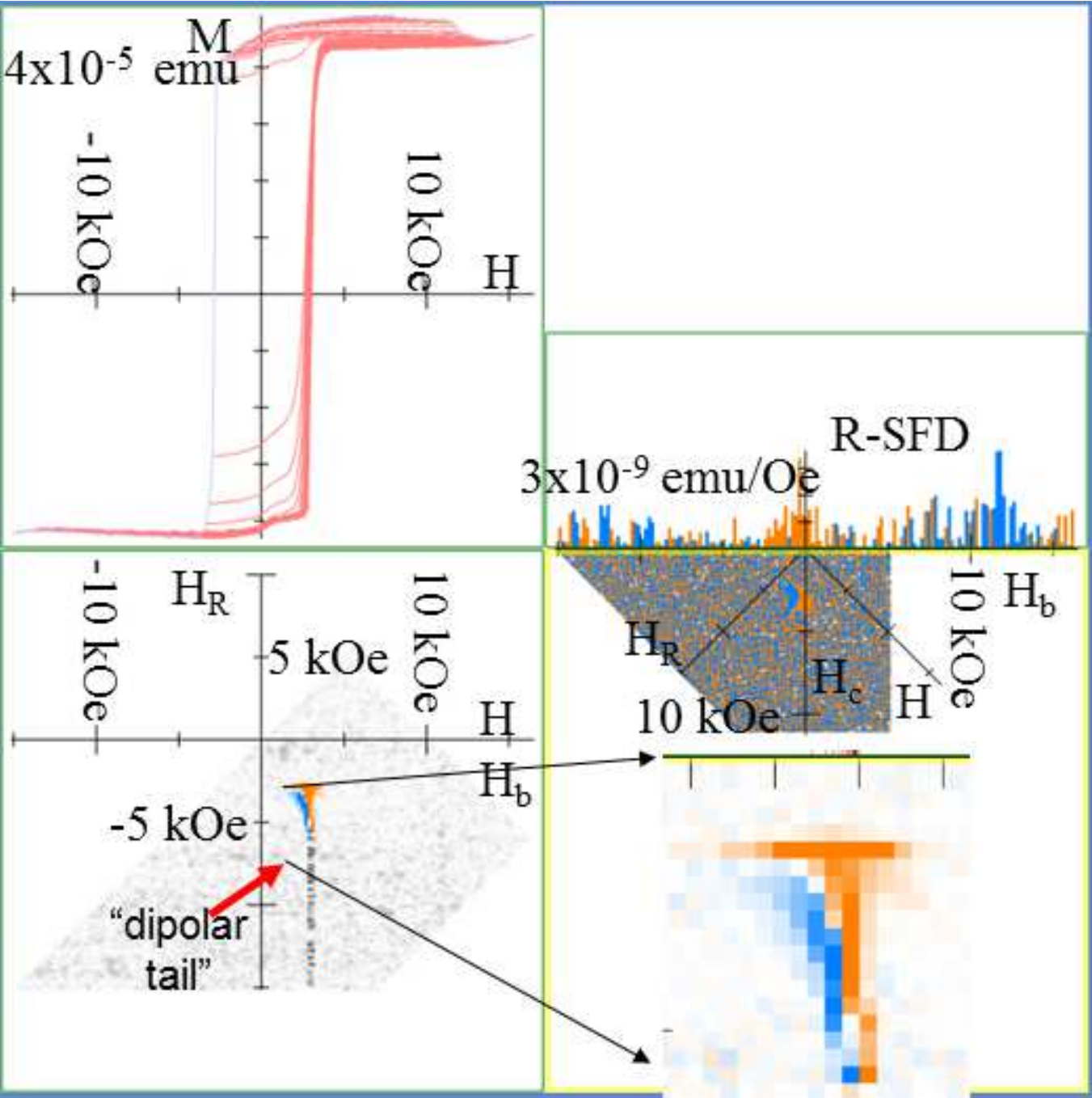}
\caption{\label{Tail} FORC+ display for an unpatterned sheet film of CoPd alloy sputtered onto a TaPd seed layer\cite{comp}, which switches by rapid domain wall motion and has very sharp features, courtesy of Dr. Bill Clark.  The insert at the lower right shows details and pixelation of the "dipole tail" that is characteristic of sudden switching.  The gain is turned up on the right side to try to see the reversible SFD, but since the reversible fraction is only 6\% in this case, it is mostly noise.}
\end{figure}

Figure \ref{Tail} shows a very different type of system, an unpatterned film that switches suddenly by domain wall motion\cite{comp}.  This leads to a very sharp feature in the FORC distribution that could be very easily obscured by smoothing -- it can be seen from the zoomed inset that the positive and negative parts of the "dipolar tail" are only one pixel wide ($\delta = 207$ Oe) at places,  This clearly demonstrates the utility of being able to view the FORC distribution without smoothing.

The FORC+ program has also been used for analysis of more complicated structures such as magnetic tunneling junctions (MTJs)\cite{NIST}, in which the characteristic fingerprints of individual layers (such as the dipole tail) can be identified in the FORC distribution of a complex structure containing such a layer.

\section{\label{concl} Conclusions}
We have presented a new method for displaying FORC data that may offer some advantages over existing display options.  It is a standalone Windows program, written in C++ and OpenGL so that it should be portable to other operating systems, and does not require any proprietary platform or visualization package.  No data manipulation is required of the user -- one simply drags a raw data file from a VSM or AGM magnetometer to the executable.  PM format files are supported in the current version, FORC+1.0, and others could easily be added. It uses a complementary-color scheme that minimizes the need for smoothing (although a polynomial-fit\cite{Egli} smoothing option is planned for a future version).  The information displayed is as close as possible to the raw data, and we can prove that no information has been lost -- the original FORC curves can be exactly recovered from the displayed information.  The avoidance of smoothing facilitates the analysis of systems with very sharp features in the FORC distribution.

\begin{acknowledgments}
We are grateful to Su Gupta, Allen Owen, Bill Clark, and Joseph Abugri for making their FORC data available and testing the FORC+ program.
\end{acknowledgments}

\nocite{*} % what does this do????

\end{document}